\documentclass{vgtc}                          

\usepackage{mathptmx}
\usepackage{graphicx}
\usepackage{times}
\usepackage{url}
\usepackage{amsmath, amsthm, amssymb}
\usepackage{color}

\definecolor{red}{rgb}{1,0,0}

\onlineid{0}

\vgtccategory{Research}

\vgtcinsertpkg

\title{Visual and semantic interpretability of projections \\of high dimensional data for classification tasks}

\author{Ilknur Icke$^1$ \thanks{e-mail: iicke@gc.cuny.edu}\\ %
      \scriptsize $^1$The Graduate Center \\ \scriptsize The City University of New York \\ \scriptsize USA
\and Andrew Rosenberg$^{1,2}$ \thanks{e-mail: andrew@cs.qc.cuny.edu}\\ %
    \scriptsize $^2$Queens College\\ \scriptsize The City University of New York \\ \scriptsize USA
}

\abstract{
 A number of visual quality measures have been introduced in visual analytics literature in order to automatically select
 the best views of high dimensional data from a large number of candidate data projections. These methods generally concentrate
 on the interpretability of the visualization and pay little attention to the interpretability of the projection axes.
 In this paper, we argue that interpretability of the visualizations and the feature transformation functions are both
 crucial for visual exploration of high dimensional labeled data. We present a two-part user study to examine these two related
 but orthogonal aspects of interpretability. We first study how humans judge the quality of 2D scatterplots of various
 datasets with varying number of classes and provide comparisons with ten automated measures, including a number of visual
 quality measures and related measures from various machine learning fields. We then investigate how the user perception on
 interpretability of mathematical expressions relate to various automated measures of complexity that can be used
 to characterize data projection functions. We conclude
 with a discussion of how automated measures of visual
 and semantic interpretability of data projections can be used together for exploratory analysis in classification tasks.

} 


\CCScatlist{
 \CCScat{H.2.8}{Database Management}{Database applications}{Data Mining}
}

\begin{document}


\firstsection{Introduction}

\maketitle


The high dimensionality of data poses theoretical and practical challenges for visual exploration and analysis.
According to the \textit{curse of dimensionality}~\cite{Bellman1961} theorem, the number of samples needed for a classification task
increases exponentially as the number of dimensions (variables, features) increases. Moreover, irrelevant and redundant
features might hinder classifier performance. On the other hand, it is costly to collect, store and process data. In exploratory
analysis settings, high dimensionality prevents the users from exploring the data visually.

Feature extraction is a two-step process that seeks suitable data representations that would help us overcome these challenges.
Feature \textit{construction} step creates a set of new features based on the original features and feature \textit{selection} is
the process of finding the best features amongst them. In this paper, we focus on feature extraction methods for visual exploration
of labeled data for classification tasks. Various linear (such as principal components analysis (PCA), multiple discriminants analysis
(MDA), exploratory projection pursuit) and non-linear (such as multidimensional scaling (MDS), manifold learning, kernel PCA,
evolutionary constructive induction) techniques have been proposed for dimensionality reduction and visualization
(~\cite{burges2005,maaten2009,guyon2006}). While each method optimizes a pre-designed intuitive measure of \textit{goodness} of
a visualization, limited number of empirical studies have been reported on how much these measures
match human perception.

In this paper, we consider two related but orthogonal aspects of human interpretability of data projections for labeled data.
Visual interpretability
is concerned with how humans judge the quality of the views presented as 2D scatterplots. In the case of labeled data,
visual interpretability closely relates to how easy it is to tell the members of each class apart by inspecting the scatterplot.
Semantic interpretability is concerned with how these views are generated, namely the complexity of the projection functions
that relate the projection axes to the original variables (features, attributes). We devised a two-part experiment to study visual
and semantic interpretability. In the first part, we showed scatterplots from four datasets with varying number of
classes and asked the participants to rate these views. The participants were not given any background information about the dataset
and the attributes since we aimed to investigate how they would rate the views independent from a specific domain. The second part of
our experiment aimed to investigate how easily humans would understand mathematical expressions with varying levels of complexity.
We generated a generic set of expressions in order to study interpretability independently from a specific domain.

This paper is organized as follows: section~\ref{sect:related} gives an overview of related work on characterization of the
interpretability of views (2D scatterplots) of labeled data. Also included in this section is an overview of related research
that addresses the interpretability of the projection (transformation) functions that generate the views.
Section~\ref{section:visualizationstudy} presents the details of our user study on how humans interpret the views of
datasets containing different number of groups and how the human perception relates to the automated measures proposed
in related literature. Section~\ref{section:mathstudy} presents our user study on how easily humans can interpret
mathematical expressions consisting of variables, coefficients and various operators and the relationship between
the human perception and automated measures of expression complexity. We conclude with a discussion (section~\ref{sect:conclusion})
of application of our results in development of visually and semantically interpretable projections of high dimensional datasets.

\section{Related Work}
\label{sect:related}

 The task of selecting interesting~\cite{friedman1974} or good views of datasets becomes more challenging as the dimensionality increases.
 For sufficiently high dimensional datasets, manual exploration of the space of views is impractical. In the case of labeled
 data, the degree of interestingness is related to how easy it is to tell the classes apart from each other by inspecting the
 visualization.

 Lee at al. present a measure for exploratory projection pursuit of labeled data that is based on
 Fisher's Linear Discriminant Analysis method in~\cite{lee2005}. The VizRank algorithm proposed by Leban et al. searches for \textit{informational} 2D projections of datasets
 that are evaluated by a k-Nearest Neighbor classifier~\cite{leban2006}. The authors claim \textit{almost perfect} agreement between
 the human judgement and the VizRank algorithm through a user study conducted using six datasets. Sips et al. propose two measures
 based on the notion of \textit{class consistency} in~\cite{sips2009}. One measure is based on preservation of closeness to class
 centroids after projection, and another is based on the entropies of the spatial distributions of the classes. The authors report
 a user study on a number of datasets with varying number of classes. They claim that their proposed measures are in alignment with
 human judgement in terms of finding all views that were labeled as good views by the participants. Tatu et al. propose two measures
 to evaluate the degree of separation on scatterplots of labeled data in~\cite{tatu2009}. Tatu et al. report a user study in~\cite{tatu2010}
 that compares four visual quality measures that have been proposed in~\cite{sips2009} and ~\cite{tatu2009}. The authors suggest
 that a combination of measures might be worth investigating.

 As opposed to the various studies on visual interpretability or quality of the projections of labeled data, the interpretability
 of the projection axes have been addressed only in a few studies. In the case of projection pursuit, the projection functions
 are given as weighted linear combinations of the original features. Morton defines the interpretability of these projection
 functions in terms of parsimony (simplicity) and proposes rotation and entropy based methods to simplify the coefficients of
 the linear projections while preserving the \textit{interesting} view~\cite{morton1989}. El-Arini et al. present a dimensionality
 reduction method that searches over scatterplots generated by simple arithmetic expressions of the original features and
 assessed by accuracy of a Bayesian classifier~\cite{elarini2006}. The authors claim that expressions containing more than
 one or two features become less interpretable. However, no empirical study has been reported to justify this intuition.

 There have been a number of related studies in the Human Computer Interaction (HCI) field investigating how humans perceive
 the complexity of mathematical expressions in order to develop improved human-computer interfaces. Anthony et al. study the
 effects of different input methods (keyboard, handwriting,
 speech) with respect to the complexity of the mathematical expressions for the
 purpose of developing intelligent tutoring systems~\cite{anthony2005} for algebra. The study presented by Awde et al.
 in~\cite{awde2008} aims to find the most appropriate way to present a mathematical expression to visually impaired users.
 The modality of the presentation is selected based on a notion of human perceived complexity of mathematical expressions
 inferred through a user study.
 Their experimental design is similar to ours, where the participants are shown a number of mathematical expressions and asked to re-write
 and rate the expressions. Then, a relationship between the structural properties of the expressions and the human perceived
 complexity is derived. The set of expressions they chose came from a wide range of fields including logic and calculus.
 In our experiments, the set of expressions were limited to various linear/non-linear combinations of
 a limited number of variables representing possible data projection functions of varying complexity.

\section{Visual Interpretability User Study}
\label{section:visualizationstudy}

 We developed computer software that automatically administered the data visualization experiment without investigator
 intervention. In this section, we present the details on design and execution of the study along with our findings on how
 well the automated measures match the human perception of visual interpretability.

\subsection{Participants}
\label{sect:participants}
 We recruited 20 participants (13 males and 7 females) who had completed or were pursuing graduate degrees in scientific
 fields such as computer science, physics, biology, engineering, accounting and psychology.

 At the beginning of the study,
 the participants were asked to fill out a brief questionnaire asking them about their related course-work or experience.
 14 of the participants specified that they had taken a Statistics, Data Mining or a Machine Learning course.

\subsection{Datasets and Visual Interpretability Measures}

  We chose four commonly used datasets from the data mining and visual analytics literature (table~\ref{tbl:datasets}).
  These datasets were selected because they contain different number of classes ranging from 2 to 9, which would let us
  investigate how the number and shape of the classes affect the relationship between human perception and the automated measures.

  The Wisconsin Diagnostic
  Breast Cancer (WDBC) dataset contains 30 measurements characterizing malignant or benign tumors. The Wine dataset contains 13
  attributes related to the chemical properties of wines from three different regions of Italy. The Segments dataset contains 19
  features derived from images of seven kinds of scenes (brickface, sky, foliage, cement, window, path, grass). All three datasets
  were downloaded from the UCI Machine Learning Repository~\cite{uci}. The Italian olive oils dataset~\cite{zupan1994} contains
  the amounts of eight fatty acids in olive oils that are from nine different regions of the country (downloaded from ~\cite{oliveoil}).

\begin{table}[h]
\begin{center}
 \scalebox{0.7}{
 \begin{tabular}{l c c c}
    \bf Name & $\#$\small \bf features & $\#$ \bf classes & $\#$  \bf scatterplots\\
 \hline
  Wisconsin Diagnostic Breast Cancer (WDBC)~\cite{uci} & 30 & 2 & 435\\
  Wine~\cite{uci} & 13 & 3 & 78\\
  Segment~\cite{uci} & 19 & 7 & 171\\
  Italian olive oils & 8 & 9 & 28\\
\hline
 \end{tabular}
 }
 \caption{{\small Datasets}}
 \label{tbl:datasets}
 \end{center}
 \end{table}

  In order to compare human judgement on quality of 2D views of labeled data to automated measures we chose ten measures from
  various fields of machine learning and visual analytics. Wrapper based methods from the feature extraction field have been
  designed to assess the usefulness of the extracted features with respect to the performance of classification algorithms.
  In this paper, we utilized four commonly used classification algorithms to assess the quality of the scatterplots for the
  four datasets mentioned above (section~\ref{sect:wrapper}). A number of cluster validity indices have been proposed in
  data clustering literature in order to assess the quality of groupings generated by different clustering algorithms. These
  indices can also be used to measure the quality of the groupings on 2D scatterplots with respect to the class labels of the
  observations. For our experiments, we included three cluster validity indices (section~\ref{sect:clustering}). A number of
  visual quality measures have been introduced in visual analytics literature (~\cite{lee2005,sips2009,tatu2009}). We included
  two of the proposed measures that were reported in~\cite{tatu2010} as the closest matches to the human perception through
  user studies (section~\ref{sect:visanalytics}). Table~\ref{tbl:visualmeasures} presents the list of the automated measures
  we utilized in our experiments.

\begin{table}[h]
\begin{center}
 \scalebox{0.8}{
 \begin{tabular}{l | l }
    \bf Name & \bf section \\
    \hline
      k-Nearest Neighbors (k-NN)~\cite{weka} & ~\ref{sect:subknn}\\
      Decision Tree (J48) ~\cite{weka} & ~\ref{sect:subdt}\\
      Naive Bayes~\cite{weka} & ~\ref{sect:subnb}\\
      Support Vector Machine (SMO)~\cite{weka} & ~\ref{sect:subsmo}\\
      \hline
      C Index ($I_{C}$) ~\cite{hubert1976}& ~\ref{sect:cindex} \\
      Davies-Bouldin Index ($I_{DB}$)~\cite{db1979} & ~\ref{sect:db} \\
      Dunn Index ($I_{Dunn}$)~\cite{dunn1974}&  ~\ref{sect:dunn}\\
      \hline
      LDA Index ($I_{LDA}$)~\cite{lee2005}& ~\ref{sect:lda}\\
      Class Consistency Measure (CCM)~\cite{sips2009} & ~\ref{sect:ccm}\\
      2D Histogram Density Measure (2D-HDM)~\cite{tatu2009} & ~\ref{sect:cdm}\\
  \hline
\end{tabular}
 }
 \caption{{\small Visual interpretability measures}}
 \label{tbl:visualmeasures}
 \end{center}
 \end{table}

 \begin{figure*}[t!]
      \begin{center}
      \includegraphics[scale=0.32]{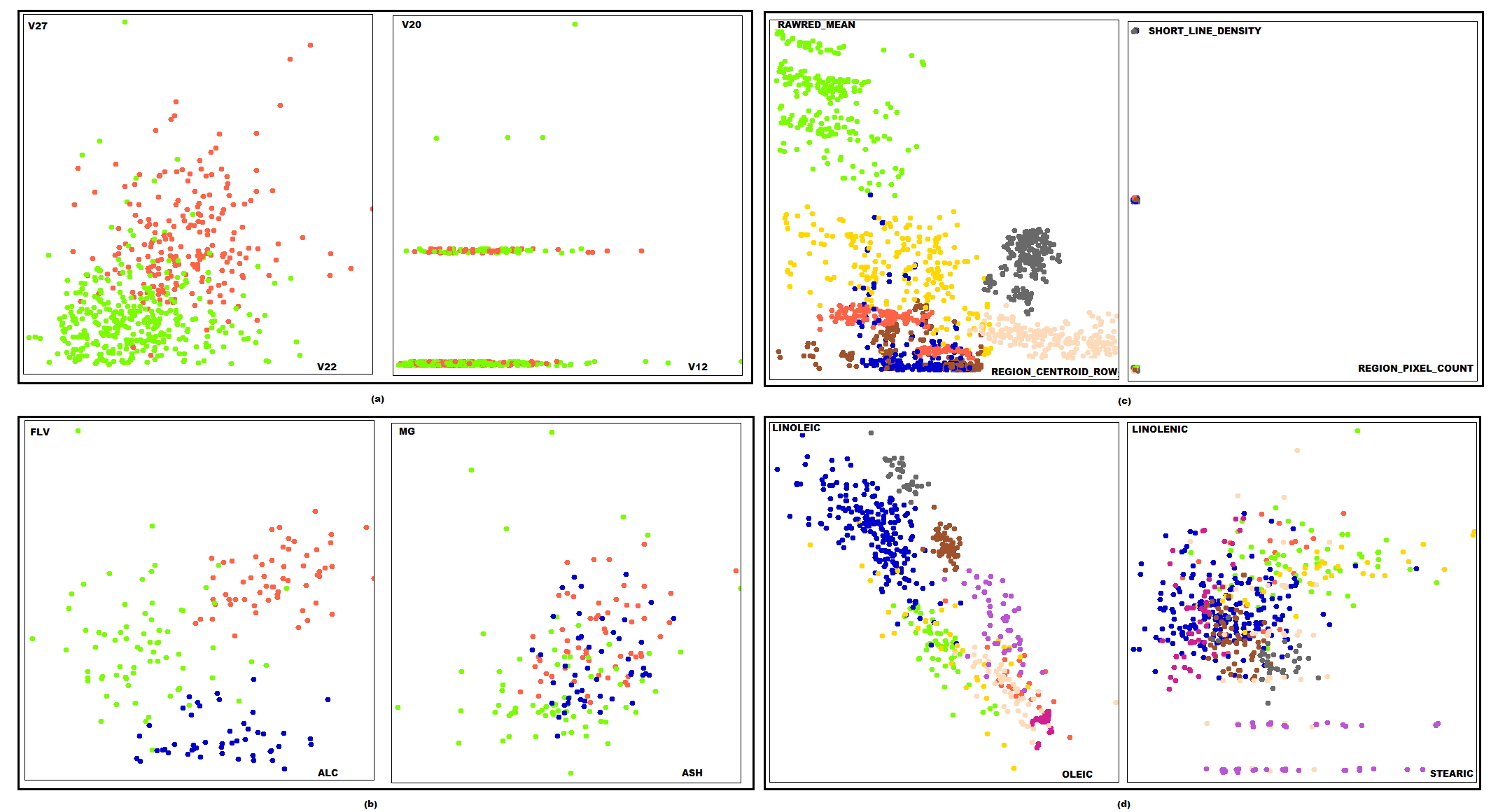}
    \caption{Best (left) and worst (right) views from the selected set of scatterplots based on the median value of all automated measures.  (a)WDBC  (b)Wine (c)Segment (d)Olive oils}
      \label{fig:oliveimages}
      \end{center}
      \end{figure*}

For a dataset with $N$ attributes, there are $N (N-1) / 2 $ unique attribute pairs, where each pair can be visualized as a
scatterplot. In order to choose the visualizations for our experiment, we first generated all possible 2D scatterplots
for each dataset.
We computed the values of the automated
measures given in table~\ref{tbl:visualmeasures} for the datasets. Our aim was to ensure that we chose a diverse set of
visualizations with respect to the automated measures. We created five equi-width bins of values between [0-1]. Then,
for each bin, we selected two scatterplots that appeared most frequently within that value range across all the automated
measures. Upon completion of this process, a total of 40 scatterplots (10 for each dataset) were selected to be included
in our experiments.

 \subsection{Wrapper Methods}

  Given a labeled multi-dimensional dataset, the goal of a supervised learning algorithm is to build a model
  from the observed data in order to predict the class membership of an unseen data item correctly. Classification algorithms can be
  broadly considered in two categories: generative and discriminative~\cite{rubinstein1997}. The generative methods aim to infer probabilistic models
  that generate the data points for each class. The discriminative methods aim to learn a mapping between the features and the
  class labels directly. Regardless of the method used, the common goal of a classification algorithm is to be able to differentiate class members as
  accurately as possible.

  Selection of appropriate features improve the performance of classifiers. Therefore, classification performance is used to
  evaluate the usefulness of feature sets in wrapper based feature selection schemes~\cite{kohavi1997}.
  Wrapper based feature selection methods can be used to assess the quality of 2D scatterplots of labeled
  data. Since the goal of a classifier is to tell the classes apart, high accuracy on two selected features
  would mean a good view of the data.

  For our experiments, we chose four of the most common (according to~\cite{wu2007}) classification algorithms which
  we briefly discuss here. Each algorithm displays a different decision boundary characteristic
  that is related to how the algorithm works.

 \begin{figure}[h!]
       \begin{center}
      \includegraphics[scale=0.28]{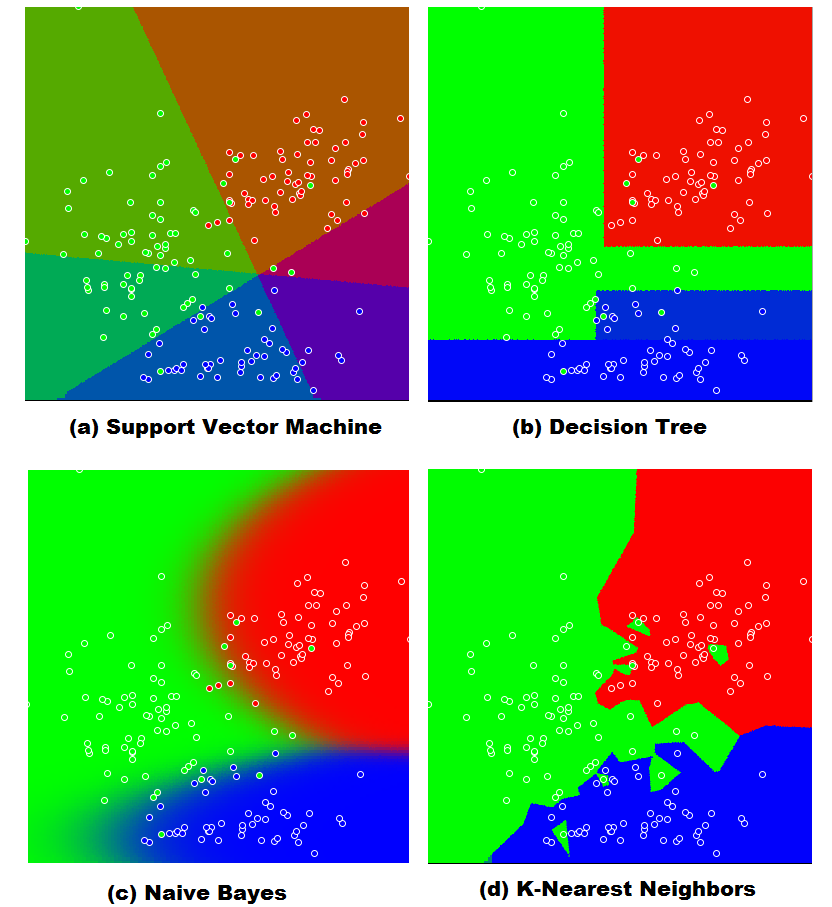}
          \caption{Classifier boundaries on a 2D view of Wine dataset}
      \label{fig:decisionboundaries}
      \end{center}
       \end{figure}

        Figure~\ref{fig:decisionboundaries} shows the decision boundaries generated by each algorithm on a
  view from the Wine dataset. The Support Vector Machine and Decision Tree algorithms generate linear decision boundaries while
  Naive Bayes and the K-Nearest Neighbors generate non-linear boundaries. Our goal is to investigate how each algorithm's decision
  boundary characteristics would relate to human perception of class separation on different datasets with varying number of
  classes.
 \label{sect:wrapper}
  \subsubsection{k-Nearest Neighbors (k-NN) Algorithm}
   \label{sect:subknn}
    In k-Nearest Neighbors algorithm, the class label of a data point is predicted based on a voting mechanism weighted by the
distances to its k closest neighbors. The distance measure is generally the Euclidean metric.
The k-Nearest Neighbors algorithm has also been utilized in Vizrank~\cite{zupan1994} in order to assess the quality of
scatterplots. In our experiments, we chose
k as $\sqrt N$ as in Vizrank,
where $N$ is the number of data points.

  \subsubsection{Decision Tree (J48) Algorithm}
  \label{sect:subdt}

   A decision tree classifier creates hierarchical partitions of the data based on one attribute at a time. The algorithm builds
   a tree structure where each internal node represents a condition that splits the dataset into multiple partitions with respect
   to a measure of partition
   impurity such as the entropy. Because of this partitioning, the decision boundaries are orthogonal to the attribute axes
   (figure ~\ref{fig:decisionboundaries}).
   In our experiments, we used the Weka implementation of the decision tree algorithm that is known as J48~\cite{weka}.

  \subsubsection{Naive Bayes Algorithm}
   \label{sect:subnb}

   The Naive Bayes algorithm is a generative classification method that estimates the joint class density separately for each class.
The assumption that all attributes are conditionally independent from each other simplifies the algorithm greatly. Despite this simplicity,
the Naive Bayes algorithm has been known to outperform more complex classification algorithms on a variety of problems~\cite{hand2001}.

  \subsubsection{Support Vector Machine (SMO) Algorithm}
   \label{sect:subsmo}

   In its simplest form, the Support Vector Machine is a machine learning technique that searches for a hyperplane
   separating the classes by the maximal margin for a two-class problem. In our experiments, we utilized the multi-class
   Weka implementation based on the Sequential Minimal Optimization (SMO) technique~\cite{weka}.

 \subsection{Cluster Validity Indices}
\label{sect:clustering}

  Data clustering is a well-known machine learning problem of categorizing multi-dimensional data into natural groupings such
  that items that are in the same group are more similar to each other than items from other groups. A number of methods have been
  proposed in order to quantify the quality of the outcome of clustering algorithms. In the case of 2D data, cluster validity indices
  can be used as measures of interpretability of the visualizations. In this section, we discuss three measures that were used in our
  experiments.
  The unifying theme of these cluster validity indices is that they all aim to measure compactness and well-separation of the class
  structures using a distance measure and they are susceptible to outliers.
  Detailed overview of validity indices can be found in~\cite{halkidi2002,rendon2011}.

  \subsubsection{C Index ($I_{C}$)}
   \label{sect:cindex}
   The C Index is a cluster validation index defined in~\cite{hubert1976}:
    \begin{equation*}
     I_{C} = \frac{SD - SD_{min}}{SD_{max}-SD_{min}}
    \end{equation*}
    where $SD$ is the total sum, for all classes, of pairwise distances between samples of the same class (total $p$ distances),
    $SD_{min}$ and $SD_{max}$ are the sums of $p$ smallest/largest pairwise distances across the whole dataset.
    The C index returns values between 0 and 1. Smaller values of $I_{C}$ indicate more compact and better separated class structures.

\subsubsection{Davies-Bouldin Index ($I_{DB}$)}
 \label{sect:db}
  The Davies-Bouldin Index is a measure of compactness and well separation of clusters and it was proposed in~\cite{db1979}:
   \begin{equation*}
     I_{DB} = \frac{1}{n} \sum_{i=1}^{n} min_{i\neq j} \{ {\frac {\delta(X_{i},X_{j})}{\Delta{(X_{i})}+\Delta{(X_{j})}}} \}
    \end{equation*}
    where $\Delta{(X_{i})}$ is intra-class distance for class $i$ and $\delta(X_{i},X_{j})$ is inter-class distance for classes $i$
    and $j$. Smaller values of $I_{DB}$ indicate more compact and better separated cluster structures. In our experiments, we
    normalized the $I_{DB}$ values to [0,1] range.

\subsubsection{Dunn's Index ($I_{Dunn}$)}
  \label{sect:dunn}
  The Dunn's index is a measure of compactness and well separation of clusters and it was proposed in~\cite{dunn1974}:
    \begin{equation*}
    I_{Dunn} = min_{1 \leq i \leq n} \{ { min_{1 \leq j \leq n, j \neq i}  \{ \frac{\delta(X_{i},X_{j})}{max_{1 \leq k \leq n} \{\Delta{(X_{k})}\}}} \} \}
    \end{equation*}
    where $\delta,\Delta$ are defined as above. Smaller values of $I_{Dunn}$ indicate more compact and better separated class
    structures. In our experiments, we normalized the $I_{Dunn}$ to [0-1] range.

\subsection{Visual Quality Measures}
\label{sect:visanalytics}
  In this section, we discuss three methods that were introduced in visual analytics literature. The Class Consistency Measure
  and the 2D Histogram Density Measure have been reported to be the closest matches to human perception amongst the four proposed
  visual quality measures through a user study~\cite{tatu2010}. Therefore, we included these measures in our experiments.

 \subsubsection{LDA Index ($I_{LDA}$)}
  \label{sect:lda}
  The LDA index is based on Fisher's discriminant analysis and has been introduced in~\cite{lee2005} for exploratory projection
  pursuit for classification problems:
    \begin{align*}
     I_{LDA} &= \frac{|W|}{|W+B|}\\
     B &= \sum^{k}_{i=1} n_{i}(\bar{V}_{i.} - \bar{V}_{..})(\bar{V}_{i.} - \bar{V}_{..})^\prime  \\
     W &= \sum_{i=1}^{k}\sum_{j=1}^{n_{i}} ({V}_{ij} - \bar{V}_{i.})({V}_{ij} - \bar{V}_{i.})^{\prime}
  \end{align*}
  where $V_{ij}$ are data points, $\bar{V_{i}}$ and $\bar{V..}$ are group and dataset centroids, k is the number of groups (classes),
  $n_{i}$ is the number of points in group $i$,  B is the between-group sum of squares and W is the within-group sum of squares.
  Smaller values of $I_{LDA}$ indicate more compact and better separated class structures.

  \subsubsection{Class Consistency Measure (CCM)}
   \label{sect:ccm}
   The Class Consistency Measure (CCM) has been proposed in~\cite{sips2009} and is based on the preservation of closeness
   to class centroid after a projection from the original data space into a 2D view.  A data point is said to be
   \textit{inconsistent}, if the projection places it closer to a class centroid other than its own. CCM scores each view with respect
   to how many consistent points it contains:

     \begin{align*}
         CCM &= 1 - \frac{\sum^{k}_{c=1} CC (x_{c})}{M} \\
         CC &= \begin{cases}
                1, \text{  if } d(x_{c},\bar{x}_{c}) < d(x_{c}, \bar{x}_{i}), 1 \leq i \leq k , i\neq c\\
                0, \text{  otherwise}
          \end{cases}
     \end{align*}
     where $M$ is the total number of data points, $k$ is the number of classes, $x_{c}$ is a point in class $c$ and
     $\bar{x}_{c},\bar{x}_{i}$ are projections of the centroid of
     classes $c$ and $i$ respectively. The CCM returns values between 0 and 1 where smaller values indicate more consistent views.

  \subsubsection{2D Histogram Density Measure (2D-HDM)}
   \label{sect:cdm}
  The Histogram Density Measure (HDM) which has been proposed in~\cite{tatu2009} is a measure of class separation based on
  2D histograms computed on 2D scatterplots of data. A weighted sum of entropy of each bin and its immediate neighbors ($u_{c}$) is
  computed as follows:

  \begin{align*}
    2D-HDM &= \frac{1}{Z} \sum_{x,y} \sum_{c} u_{c} (-  \sum_{c} \frac{u_{c}} { \sum_{c} {u_c}} log_{2} \frac{u_{c}}{\sum_{c}{u_c}}    ) \\
    \frac{1}{Z} &= \frac{1} {log_{2}M \sum_{x,y} \sum_{c} u_{c}}
  \end{align*}
   where $\frac{1}{Z}$ is a normalization factor in order to confine the values within the [0-1] range and smaller values indicate
 better data separation. The choice of bin size influences how this measure scores the views. In our experiments, we used 100x100 bins
 for each dataset.

\subsection{Task}

  Before starting the study, the participants were told that they would be shown a series of scatterplots of some datasets
  containing multiple groups and their task was to rate how good the view was by inspecting the visualizations. We did not
  provide any directions on how to define the ``goodness" of a view. Each scatterplot showed only the data points in different
  colors with respect to their class labels and no further information about the data (such as the name of the dataset or the
  names of the attributes) was provided. After reading the instructions, the participants  were shown one scatterplot at a time and
  were asked to rate them on a continuous scale between 0 (very good)  to 1 (very bad) with labels shown on table~\ref{tbl:visRating}.

\begin{table}[h]
\begin{center}
 \scalebox{0.7}{
 \begin{tabular}{l | c}
    \bf Rating & \bf Value\\
    \hline
    Very Good & 0.0 \\
    Good & 0.25 \\
    Average & 0.5 \\
    Bad & 0.75 \\
    Very Bad & 1.0 \\
\hline
 \end{tabular}
 }
 \caption{{\small Labels for visual interpretability ratings on continuous scale}}
 \label{tbl:visRating}
 \end{center}
 \end{table}

\subsection{Methodology}

  Each user rated a total of 45 scatterplots. Undisclosed to the participants, the first five scatterplots were artificial views
  showing different levels of compactness and separation between the classes from \textit{very good} to \textit{very bad}
  with respect to the automated measures. These visualizations were used as \textit{calibration views} in order to help
  the participant get used to the interface and build their mental models for how they would rate the quality of a view.
  The user ratings for these calibration views were not included in the analysis of the responses. The remaining 40 pre-selected
  scatterplots were displayed in a randomized order to each user. In order to reduce the effect of outliers, we computed the
  median of user responses for each of the scatterplots and used this value in our comparisons to the automated measures.

\subsection{Results}

\begin{figure*}[t!]
      \begin{center}
      \includegraphics[scale=0.28]{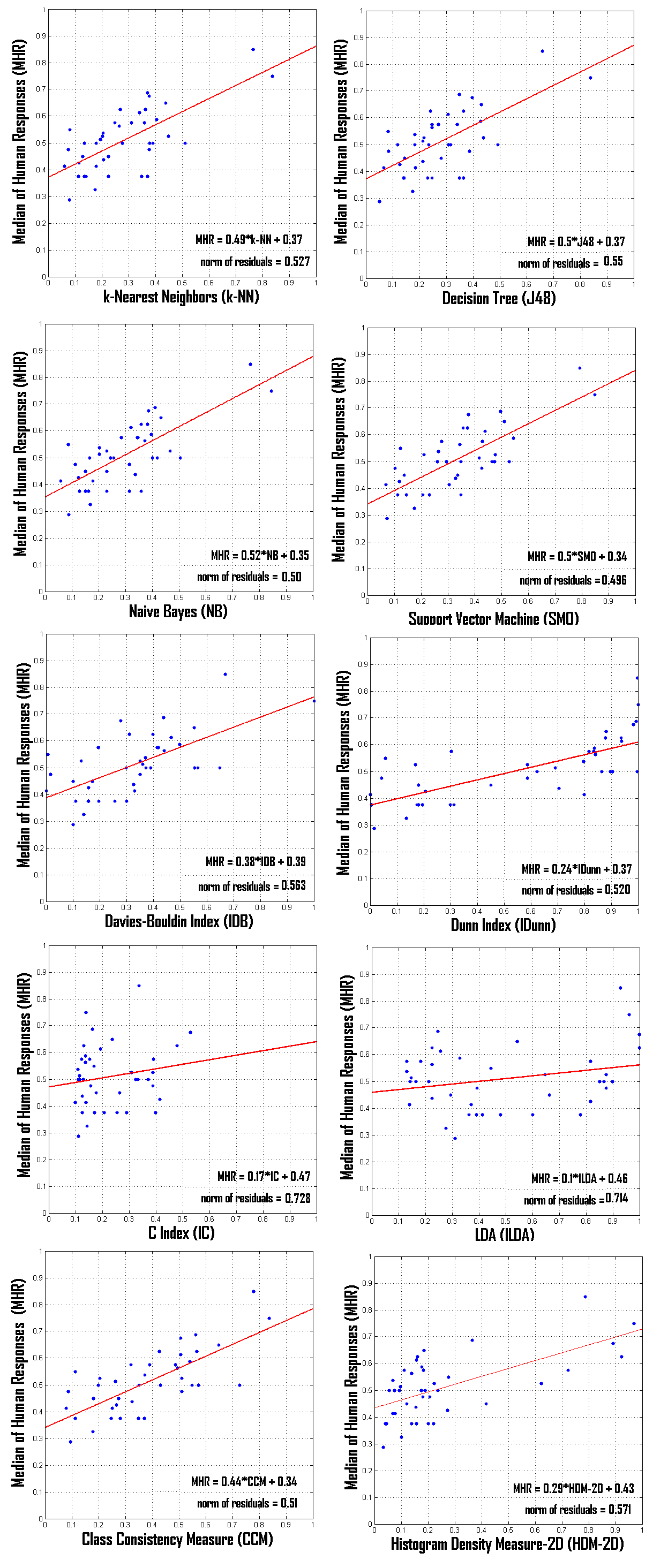}
    \caption{\small Human responses versus the automated measures on all scatterplots}
      \label{fig:allmeasures}
      \end{center}
      \end{figure*}

     Figure~\ref{fig:allmeasures} shows the relationships between the human perception and
     each of the automated measures. Each plot presents the values of the corresponding automated
     measure versus the median value of the participant ratings for each scatterplot. A strong positive linear correlation
     means good alignment between the human and the automated measure.

  Table~\ref{tbl:visresults1} summarizes the relationships
     between each measure and human perception for all scatterplots. Tables~\ref{tbl:visresults2}-~\ref{tbl:visresults5}
     present the results for each individual dataset. The measures are sorted in descending order with respect to
     the $R^{2}$.

\begin{table}[h!]
\begin{center}
 \scalebox{0.7}{
\begin{tabular}{l | c | c | c | c | c}
   \bf Measure & \bf SSE & \bf $R^{2}$ & \bf Adj. $R^{2}$ & \bf RMSE & \bf p-value \\
   \hline
    Support Vector Machine &  0.2461 & 0.5506 & 0.5388 & 0.0805 & $<$ 0.05 \\
    Naive Bayes &  0.2516 & 0.5406 & 0.5285 & 0.0814 & $<$ 0.05  \\
    Class Consistency Measure & 0.2603 & 0.5246 & 0.5121 & 0.0828 &$<$ 0.05\\
    Dunn Index & 0.2712 & 0.5047 & 0.4916 & 0.0845& $<$ 0.05\\
    K-Nearest Neighbors & 0.2785 & 0.4914 & 0.4780 & 0.0856 & $<$ 0.05 \\
    Decision Tree & 0.3064 & 0.4404 & 0.4257 & 0.0898 &$<$ 0.05 \\
    Davies-Bouldin Index & 0.3178 & 0.4197 & 0.4044 & 0.0914& $<$ 0.05\\
    2D-Histogram Density Measure & 0.3258 & 0.4050 & 0.3893 &0.0926&$<$ 0.05\\
    LDA Index & 0.5112 & 0.0664 & 0.0419 &0.1160& 0.11\\
    C Index & 0.5314 & 0.0295 & 0.0040 & 0.1183&0.29\\
 \hline
 \end{tabular}
  }
 \caption{{\small Summary of linear relationships (Df:38, $\alpha=0.05$) between the automated measures and human perception on all scatterplots}}
 \label{tbl:visresults1}
  \end{center}
 \end{table}

\begin{table}[h!]
\begin{center}
 \scalebox{0.7}{
\begin{tabular}{l | c | c | c | c | c }
   \bf Measure & \bf SSE & \bf $R^{2}$ & \bf Adj. $R^{2}$ & \bf RMSE & \bf p-value\\
    \hline
    2D-Histogram Density Measure & 0.0208 & 0.7592 &0.7291&0.0510 &$<$ 0.05  \\
    Support Vector Machine & 0.0216 & 0.7495 &0.7182 &0.0520 &$<$ 0.05 \\
    K-Nearest Neighbors & 0.0277 &  0.6786 & 0.6385 &0.0589& $<$ 0.05\\
    Naive Bayes & 0.0280 & 0.6759 & 0.6353 & 0.0591&$<$ 0.05\\
    Decision Tree & 0.0288 & 0.6667&0.6250&0.0600&$<$ 0.05\\
    Dunn-Index & 0.0340 & 0.6057 & 0.5564 & 0.0652& $<$ 0.05\\
    Class Consistency Measure & 0.0481&0.4430&0.3733&0.0775 & $<$ 0.05\\
    Davies-Bouldin Index & 0.0559 & 0.3516 & 0.2706&0.0836 & 0.07\\
    LDA Index & 0.0637 & 0.2616 & 0.1693 &0.0892& 0.13\\
    C Index& 0.0661 &0.2337&0.1379 &0.0909&0.16\\
 \hline
 \end{tabular}
  }
 \caption{{\small Summary of linear relationships (Df:8, $\alpha=0.05$) between the automated measures and human perception for WDBC dataset (2 classes)}}
 \label{tbl:visresults2}
  \end{center}
 \end{table}

 \begin{table}[h!]
\begin{center}
 \scalebox{0.7}{
\begin{tabular}{l | c | c | c | c | c}
   \bf Measure & \bf SSE & \bf $R^{2}$ & \bf Adj. $R^{2}$ & \bf RMSE & \bf p-value\\
    \hline
    Dunn Index & 0.0087 & 0.8061 & 0.7818 & 0.0329 &  $<$ 0.05 \\
    Support Vector Machine & 0.0118 &  0.7363 &0.7033 &0.0384 & $<$ 0.05\\
    Class Consistency Measure &0.0145 & 0.6764 &0.6359 &0.0426 & $<$ 0.05\\
    Decision Tree & 0.0155 & 0.6535 &  0.6102 & 0.0440 &$<$ 0.05 \\
    C Index & 0.0167 & 0.6266 & 0.5799 & 0.0457 & $<$ 0.05\\
    Naive Bayes &  0.0190 & 0.5761 & 0.5231 &0.0487&$<$ 0.05 \\
    LDA Index & 0.0191& 0.5741&0.5209&0.0488&$<$ 0.05\\
    Davies-Bouldin Index & 0.0201& 0.5519& 0.4959&0.0501&$<$ 0.05\\
    K-Nearest Neighbors & 0.0208 & 0.5357 &0.4777&0.0510&$<$ 0.05\\
    2D-Histogram Density Measure&0.0245&0.4531&0.3847&0.0553&$<$ 0.05\\
  \hline
 \end{tabular}
  }
 \caption{{\small Summary of linear relationships (Df:8,  $\alpha=0.05$) between the automated measures and human perception for Wine dataset (3 classes)}}
 \label{tbl:visresults3}
  \end{center}
 \end{table}

\begin{table}[h!]
\begin{center}
 \scalebox{0.7}{
\begin{tabular}{l | c | c | c | c | c}
   \bf Measure & \bf SSE & \bf $R^{2}$ & \bf Adj. $R^{2}$ & \bf RMSE & \bf p-value\\
    \hline
  2D-Histogram Density Measure&0.0459&0.7140&0.6783&0.0757 & $<$ 0.05 \\
  Dunn Index             & 0.0617 & 0.6152   & 0.5671 & 0.0878&$<$ 0.05\\
  Naive Bayes            & 0.0635 & 0.6041   & 0.5546 & 0.0891&$<$ 0.05\\
  Support Vector Machine & 0.0636 & 0.6035   & 0.5539 & 0.0891&$<$ 0.05\\
  K-Nearest Neighbors    & 0.0708 & 0.5582   & 0.5030 & 0.0941&$<$ 0.05\\
  Decision Tree          & 0.0766 & 0.5222   & 0.4625 & 0.0979&$<$ 0.05\\
  Class Consistency Measure &0.0906 & 0.4352 & 0.3647 & 0.1064&$<$ 0.05\\
  Davies-Bouldin Index   &0.1134 & 0.2927 & 0.2043 & 0.1191 & 0.11\\
  LDA Index              & 0.1467 & 0.0847 & -0.0297 & 0.1354&0.41\\
  C Index                & 0.1603 & 0.0002  & -0.1248 & 0.1416&0.98\\

 \hline
 \end{tabular}
  }
 \caption{{\small Summary of linear relationships (Df:8, $\alpha=0.05$) between the automated measures and human perception for Segment dataset (7 classes)}}
 \label{tbl:visresults4}
  \end{center}
 \end{table}

 \begin{table}[h!]
\begin{center}
 \scalebox{0.7}{
\begin{tabular}{l | c | c | c | c | c}
   \bf Measure & \bf SSE & \bf $R^{2}$ & \bf Adj. $R^{2}$ & \bf RMSE & \bf p-value\\
   \hline
   Naive Bayes & 0.0108& 0.7420&0.7097&0.0368&$<$ 0.05\\
   Davies-Bouldin Index &0.0109 &0.7405&0.7080& 0.0369&$<$ 0.05\\
   K-Nearest Neighbors & 0.0114&0.7275&0.6935&0.0378&$<$ 0.05\\
   Class Consistency Measure & 0.0122 &  0.7100  & 0.6738 & 0.0390&$<$ 0.05\\
   C Index & 0.0138&0.6702&0.6289&0.0416&$<$ 0.05\\
   2D-Histogram Density Measure&0.0147&0.6496&0.6058&0.0428&$<$ 0.05\\
   Decision Tree & 0.0149 & 0.6445 &0.6001&0.0431&$<$ 0.05\\
   Support Vector Machine & 0.0188 &0.5513 & 0.4952&0.0485&$<$ 0.05\\
   LDA Index & 0.0212 &0.4950&0.4319&0.0514&$<$ 0.05\\
   Dunn Index & 0.0367 &0.1252&0.1252&0.0677& 0.32\\
    \hline
 \end{tabular}
  }
 \caption{{\small Summary of linear relationships (Df:8, $\alpha=0.05$) between the automated measures and human perception for Olive Oils dataset (9 classes)}}
 \label{tbl:visresults5}
  \end{center}
 \end{table}

  The $R^{2}$ values indicate how well the linear regression model fits the data in terms of predicting the participant ratings
  based on each automated measure. Results on all scatterplots  without consideration of a specific
  dataset (table~\ref{tbl:visresults1}) show that the C Index and the LDA Index did not correlate
  with how the participants typically judged the ``goodness" of a view. Two classification algorithms, Support Vector Machine
  and Naive Bayes ranked the top two, tightly followed by the Class Consistency Measure, Dunn Index and K-Nearest Neighbors.

  When we look at the results on individual datasets (tables~\ref{tbl:visresults2}-~\ref{tbl:visresults5}), we notice that the
  Support Vector Machine is no longer one of the top performers for datasets with larger number of classes (Segments and Olive oil).
  The Dunn Index shows a significant match for all datasets accept for the Olive oils dataset which might be due to the fact
  that, the class structures contain outliers and in some cases, the members of the same class are clumped together
  in multiple areas on the scatterplot.  Overall, we found that Davies-Bouldin Index, Dunn Index, C Index and LDA Index might
  not correlate with human perception, depending on the dataset while all others seem to correlate to some extent.

  From these results, we infer that the degree of match between the human and the automated measure might depend on
  the characteristics of the views such as the shape of the clusters formed by the class members. Therefore, we hypothesize
  that a combination
  of these measures might model the
  human perception better than any single measure. In the next section, we derive a composite measure based on the
  individual measures and investigate its performance across all individual datasets included in our experiments.

\subsection{Combining the Automated Visual Interpretability Measures}

\begin{figure}[h!]
       \begin{center}
      \includegraphics[scale=0.37]{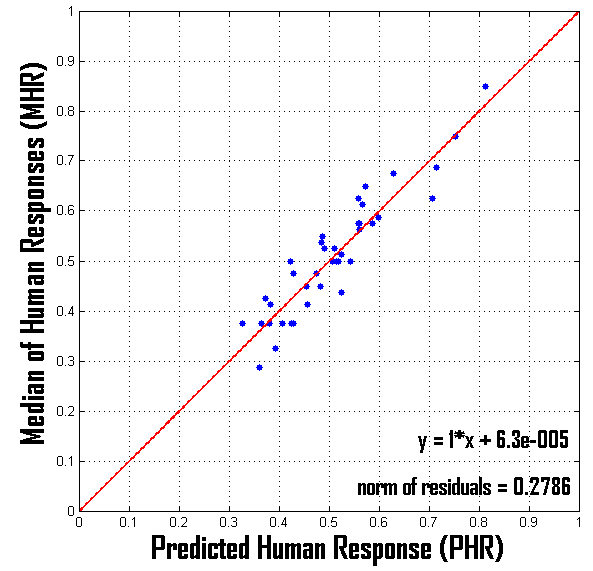}
          \caption{Human responses versus the inferred aggregate model of automated measures on all scatterplots}
      \label{fig:combinedallviews}
      \end{center}
       \end{figure}

    Given the ten automated measures and the median of human responses for all datasets, we cast a prediction problem that would learn a linear model for the human responses in terms of the automated
    measures. We trained a linear regression model with leave-one-out cross-validation. The following linear combination of six of the
    ten measures was found (figure~\ref{fig:combinedallviews}):

   \begin{align*}
    Predicted Human Response (PHR) =& -0.7772 * J48 +\\
      &0.8155 *  SMO +\\
     &-0.4305 *   I_{C} +\\
     &-0.4588 *   I_{DB} +\\
     & 0.6586 *  CCM  +\\
     & 0.3285 *  HDM-2D +\\
     & 0.3606
   \end{align*}


    As it can be seen on table~\ref{tbl:combinedmeasure}, on the set of all scatterplots,
    the combined measure matches the human perception significantly better than any single measure reported on
    table~\ref{tbl:visresults1}. The composite measure is the clear winner for the WDBC and the Segment datasets. However,
    for Wine and Olive oil datasets, it was not the top measure in terms of matching human perception.
    Overall, the correlation between the composite measure and human perception was significant across all scatterplots as well
    as individual datasets.

    One interesting observation is that, although
    the two classification algorithms that are related to density estimation (Naive Bayes and k-Nearest Neighbours) were always
    in alignment with human perception on all datasets, they were excluded from the composite model in favor of the discriminative
    classifiers.

\begin{table}[h!]
\begin{center}
 \scalebox{0.7}{
\begin{tabular}{l | c | c | c | c| c | c}
   \bf Dataset & \bf SSE & \bf $R^{2}$ & \bf Adj. $R^{2}$ & \bf RMSE &\bf Df & \bf p-value \\
   \hline
   All  & 0.07764 & 0.8582 & 0.8545&0.0452 & 38 & $<$ 0.05\\
   WDBC & 0.0118 & 0.8629&0.8457 & 0.0385& 8 & $<$ 0.05\\
   Wine & 0.0164 & 0.6331&0.5872 & 0.0453 & 8 &$<$ 0.05 \\
   Segment &0.0138&0.9139& 0.9031 & 0.0415 & 8 & $<$ 0.05\\
   Olive oils & 0.018 & 0.5702  & 0.5165 &0.0474 & 8 & $<$ 0.05\\
 \hline
 \end{tabular}
  }
 \caption{{\small The degree of match between the combined linear model of the automated measures and
           human perception on combined and individual datasets ($\alpha = 0.05$)}}
 \label{tbl:combinedmeasure}
  \end{center}
 \end{table}

\section{Semantic Interpretability User Study}
\label{section:mathstudy}

 The goal of the semantic interpretability study is to understand how easily the users interpret/understand mathematical
 expressions of variables, coefficients and operators that would make up a linear or non-linear projection (transformation)
 function characterizing the relationship between a set of variables and result of the projection.

  We developed a software
 application that automatically administered the experiment and recorded participant responses without investigator intervention.
 The participants used a consumer grade tablet pen interface connected to the computer via a USB connection.

\subsection{Participants}
  The same 20 participants (section~\ref{sect:participants}) that were involved in the visualization study also took part
  in this experiment. Before starting the study, all participants were given time to train on using the tablet pen interface.

\subsection{The expressions}
  We created 30 mathematical expressions consisting of five possible variables $t,u,x,y,z$, numerical coefficients,
  mathematical operators $+,-,*,/$, logarithm, square-root, exponential and power. Undisclosed to the participants,
  the first five expressions were used as \textit{calibration expressions} (table~\ref{tbl:calibrationexpressions}).
  The purpose of these initial expressions was to establish a range of complexity of the expressions that would be shown further on.
  The participant responses to these expressions were not included in analysis of the results. In our experiments, the size of
  the shortest expression was 2 and the longest expression was 19.

\begin{table}[h]
\begin{center}
 \scalebox{0.67}{
\begin{tabular}{l | c | c | c | c | c }
                & \bf $\#$    &\bf $\#$      & \bf Tree               &\bf $\#$  & \bf Total \\
 \bf Expression & \bf Operands & \bf Operators &  \bf Depth & \bf Blocks & \bf Size \\
 \hline
  $x^{2}$ & 1 & 1 & 2 & 1 & 2\\
  $x * y + z * u$ & 4 & 3 & 3 & 2 & 7\\
  $2*log(z) + \sqrt{x}$ & 3 & 4 & 4 &2 & 7 \\
  $e^{\sqrt{t+u}  \textbf{ } / \textbf{ } ( \textbf{ } log(x) \textbf{ } * \textbf{ } ( \textbf{ } log(u)+log(z) \textbf{ } ) \textbf{ } ) }$ & 5 & 9 & 6 & 3 & 14\\
  $( \textbf{ } 0.5 * t * \sqrt{(u*y) + t} \textbf{ } ) \textbf{ } / \textbf{ } ( e^{\sqrt{x*log(y)} \textbf{ } / \textbf{ } z} )$ & 8 & 11 & 7 & 3 & 19 \\
\hline
\end{tabular}
  }
 \caption{{\small Five mathematical expressions used as calibration stimuli}}
 \label{tbl:calibrationexpressions}
  \end{center}
 \end{table}

\begin{table*}[t!]\renewcommand{\arraystretch}{1.2}\addtolength{\tabcolsep}{-1pt}
\begin{center}
 \scalebox{0.75}{
\begin{tabular}{l | c | c | c | c | c |c || c |c | c}
 \bf Expression& \bf $\#$     & \bf $\#$              & \bf Tree     & \bf $\#$     & \bf Avg. & \bf Total & \bf Median of  &\bf Median of  & \bf $\#$ Correct \\
                & \bf Operands &  \bf Operators        & \bf Depth    &  \bf Blocks & \bf Block &  \bf Size & \bf Human &\bf Time Spent &  \bf (out of 20)  \\
                &              &                       &              &             & \bf Size &          & \bf Ratings   &\bf Writing (seconds) &  \\
 \hline
 $log(x)$ & 1 & 1 & 2 & 1 & 2 & 2 & 0.0 & 18.69 & 20\\
 $0.5 * t$ & 2 & 1 & 2 & 1 & 3 & 3 & 0.025 & 17.55 & 20 \\
 $x \textbf{ }/ \textbf{ } (y - 1)$ & 3 & 2 & 3 & 2 & 2 &  5&  0.125 & 19.6 & 18\\
 $\sqrt{ log (y)} \textbf{ } / \textbf{ } y$& 2 & 3 & 4& 2& 2& 5 & 0.3 & 21.1 & 16\\
 $log (t) + log (u)$ &2& 3& 2 & 2& 2 &5 & 0.225&21.15&20\\
 $e^{\sqrt {log ( t^{2} )} }$ &  1 & 4 & 5 & 1 & 5 &5&0.25&20.37&19\\
 $u * ( z + x)$ & 3 & 2 & 3 & 2 & 2 &5 & 0.2&19.08&19 \\
 $t + ( x * z )$ & 3 & 2 & 3 & 2 & 2 & 5 & 0.25&23.19&19\\
 $0.2*t + u * y$ & 4 & 3 & 3 & 2& 3& 7 & 0.25&23.48&20\\
 $e^{(t + u) \textbf { } * \textbf { } (z-1)}$ & 4 & 4 & 4 & 2 & 3 &8 & 0.475&17.46&13\\
 $( t  + u + x ) \textbf{ }*\textbf{ } ( y - z)$ & 5 & 4 & 4 & 2 & 4 &9 & 0.375&20.61&13\\
 $( \textbf{ }( x * y ) - z \textbf{ }) \textbf{ }/ \textbf{ }( t - u )$ & 5 & 4 & 4 &2 & 4&9 & 0.5&24.24&11\\
 ${ ( \sqrt{t}  + (z*x)  + \sqrt{u} )} ^2$ & 4 & 6 & 5 & 3 & 2.33& 10 & 0.56&17.47&11\\
 $y \textbf{ } / \textbf{ } (  \textbf{ } ( \textbf{ }( t^{2} + u ) * z \textbf{ })  + x \textbf{ })$ &5 & 5 & 6 & 3 &2.67 &10 & 0.612& 19.83&6\\
 $t \textbf{ } / \textbf{ } (  \textbf{ } ( \textbf{ }( t^{2} - t ) * u \textbf{ })  - u \textbf{ })$ & 5 & 5 & 6 & 3 & 2.67&10 & 0.625&22.45&10\\
 $( \textbf{ }(t * x) / u\textbf{ })   \textbf{ } - \textbf{ } ( \textbf{ }( y - z ) / x \textbf{ })$& 6 & 5 & 4 & 2 &5 &11 & 0.625&17.44&2\\
 $log (x ) + log (y) - log(t) - log(x)$ & 4 & 7 & 5 & 4 & 2 &11&0.375&15.39&12\\
 $(z / x ) \textbf{ } + \textbf{ } ( \textbf{ }(x / y ) \textbf{ } / \textbf{ } (t / u ) \textbf{ })$ & 6& 5 & 4 & 3 &3&11&0.625&26.05&7\\
 $e^{ (\sqrt{t} + \sqrt{u}  ) \textbf{ }*\textbf{ }  ( x + (y * z) ) }$ & 5 & 7 & 5 & 2 & 5 &12 &0.712&21.85&7\\
 $log ( \textbf{ }(\sqrt{t} * \sqrt {u} )  \textbf{ }+\textbf{ }  ( x * (t + u)\textbf{ } ) \textbf{ })$ & 5 & 7 & 5 & 2 & 5&12 & 0.75&22.64&6\\
 $((x-1)*(y-2))  \textbf{ } / \textbf{ } (t / u)^2$ & 6 & 6 & 4 & 3 & 3.33&12 & 0.575&19.07&10\\
 $(\textbf{ } (log(t) * \sqrt{u})  + (t + x) - y^2\textbf{ } ) \textbf{ } / \textbf{ }  (t * z)$ & 7 & 9 & 5 & 3 & 4.67& 16 & 0.85&18.41&0\\
 $( \textbf{ }( 0.2*e^{-2*x})  \textbf{ } / \textbf{ }  (z + x) \textbf{ }) -  (\sqrt{y} / (u * x) \textbf{ })$ & 8 & 9 & 6 & 3 & 5& 17 & 0.875&19.27&0 \\
 $(t * x) + (u * y )  + (z * y) + (x * u) + (t * z)$ & 10 & 9 & 5 & 5 & 3&19 &0.637&19.63&2\\
 $0.1 * t + 0.5 * u + 0.2 * z + 0.4 * y + 0.3 * x$ & 10 & 9 & 5 & 5 & 3 &19 & 0.75&22.16&1 \\

 \hline
\end{tabular}
  }
 \caption{{\small 25 mathematical expressions used in assessment of expression interpretability (ordered by total size)}}
 \label{tbl:assessmentexpressions}
  \end{center}
 \end{table*}

\subsection{Task}
 The participants were informed that they would be shown a series of mathematical expressions of five possible variables
 $t,u,x,y,z$, numerical coefficients, mathematical operators $+,-,*,/$, logarithm, square-root, exponential and power. They
 were told that each expression would be displayed for 10 seconds and that their task was to study the expression within
 that time and write it back using the tablet pen after the expression was removed from the screen. They were also asked
 to rate how easy it was to understand/interpret the given expression. The rating was on a continuous scale from 0 (very easy)
 to 1 (very difficult) with labels shown on table~\ref{tbl:mathratings}.

\begin{table}[h]
\begin{center}
 \scalebox{0.7}{
 \begin{tabular}{l | c}
    \bf Rating & \bf Value\\
    \hline
    Very Easy & 0.0 \\
    Easy & 0.25 \\
    Average & 0.5 \\
    Difficult & 0.75 \\
    Very Difficult & 1.0 \\
\hline
 \end{tabular}
 }
 \caption{{\small Labels for semantic interpretability ratings on continuous scale}}
 \label{tbl:mathratings}
 \end{center}
 \end{table}

\subsection{Methodology}

  After the first five calibration expressions, the remaining 25 expressions (table~\ref{tbl:assessmentexpressions})
  were then shown in a randomized order to each participant. The expressions were presented to the participants
  in a linearized form. Specifically, we chose not to display the division operation as a fraction in order not to create
  a visual cue that would make it easier to interpret the expression as opposed to addition, subtraction or multiplication.

\begin{figure*}[t!]
      \begin{center}
      \includegraphics[scale=0.3]{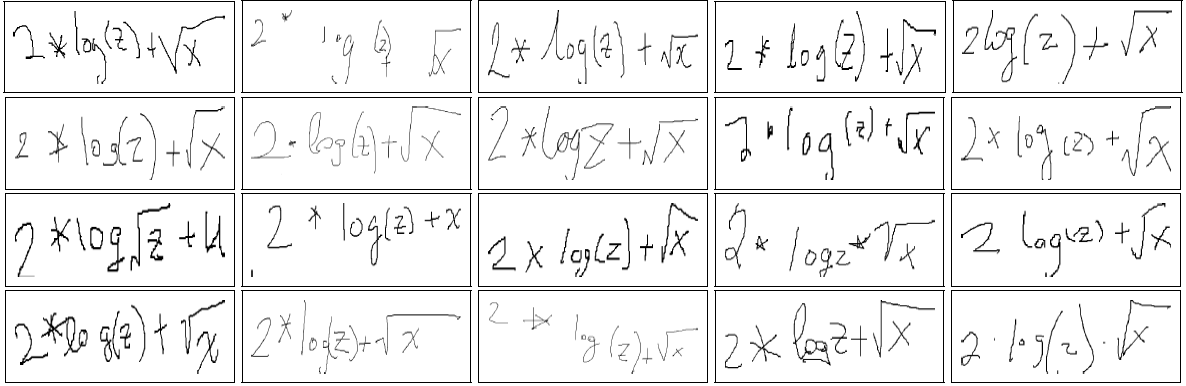}
      \caption{A sample of participants' entries using tablet pen interface (expression: $2*log(z) + \sqrt{x}$ )}
      \label{fig:handwriting}
      \end{center}
      \end{figure*}

  For each participant, we recorded the time they spent writing each expression and their rating on how easy it was
  to understand the expression. The images of the expressions they wrote down were automatically captured and saved
  to disk for manual inspection for correctness (figure~\ref{fig:handwriting}). For this study, we only considered correct/incorrect
  response rather than assessing partial correctness.

     \begin{figure}[h]
      \begin{center}
      \includegraphics[scale=0.5]{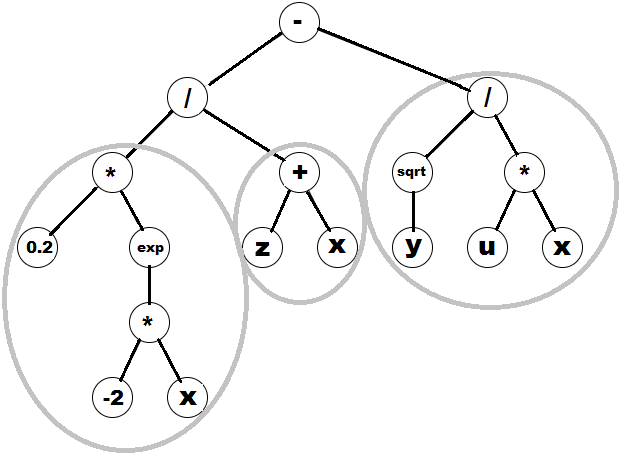}
    \caption{Expression tree for $( \textbf{ }( 0.2*e^{-2*x})  \textbf{ } / \textbf{ }  (z + x) \textbf{ }) -  (\sqrt{y} / (u * x) \textbf{ })$. Tree depth:6,
    $\#$ of blocks:3, $\#$operands:8, $\#$operators:9}
      \label{fig:expressiontree}
      \end{center}
      \end{figure}

\subsection{Results}

    The outcome of the study is summarized on table~\ref{tbl:assessmentexpressions}. For each expression, we report the median
    value of the participant ratings, median value of the total time it took for the participants to write down and rate the
    expression and the number of correct responses. We first examined the relationship between how an expression is rated
    by the participants and how frequently it was written down correctly. We hypothesized that the expressions that were frequently
    replicated incorrectly would also be rated as difficult to interpret by the participants. Indeed, we found that there was
    a strong
    correlation between them (Pearson's R=$0.9379$, Df=23, $p < 0.05$) indicating that the ratings given by the participants were
    consistent with their observed behavior (answering correctly/incorrectly). We did not find a meaningful relationship between
    the time taken to write the expression and the subjective rating. Most possibly, it is due to the fact that the participants
    did not spend much time on those expressions that they found hard to interpret. Therefore, we will present only the results
    based on the participant ratings here.

    In order to define a relationship between the structure of the expression and the participant ratings, we determined various
    attributes of the expressions characterizing the complexity. For this purpose, we utilized a syntax-tree representation for
    the expressions (figure~\ref{fig:expressiontree}). The number of operands include the variables and other numerical values.
    A block is a sub-expression composed of multiple operators and operands. The tree depth relates to the nestedness in
    the expression and the number and size of the blocks indicate
    the distinguishable components it contains. Table~\ref{tbl:assessmentexpressions} shows the values of the six derived
    attributes for each expression.

     Table~\ref{tbl:assessmentexpressions} reveals that up to expression size 7 (up to 4 operands or 4 operators),
    the participants rated the expressions to be very easy-easy (rating $\leq 0.3$).

     We first looked at how much each of these attributes can predict the participant's ratings on interpretability
     (table~\ref{tbl:expressionAttributes}). Unsurprisingly, the number of operators and total size affect how the
     expressions are rated. But this does not explain why the expressions of the same size were rated differently by
     the participants.

 \begin{table}[h!]
 \begin{center}
 \scalebox{0.7}{
\begin{tabular}{l | c | c | c | c |c }
   \bf Expression Attribute & \bf SSE & \bf $R^{2}$ & \bf Adj. $R^{2}$ & \bf RMSE & \bf p-value\\
   \hline
   Number of Operators & 0.3015 & 0.8023 &0.7937 &0.1145 & $ < 0.05$\\
   Total Size          & 0.3130 & 0.7948 &0.7859 &0.1167 & $ < 0.05$\\
   Number of Operands  & 0.5177 & 0.6606 &0.6458 &0.1500 & $ < 0.05$\\
   Tree Depth          & 0.5290 & 0.6532 &0.6381 &0.1517 & $ < 0.05$\\
   Number of Blocks    & 0.9704 & 0.3638 &0.3361 &0.2054 & $ < 0.05$\\
   Avg. Block Size     & 1.0060 & 0.3406 &0.3119 &0.2091 & $ < 0.05$\\
   \hline
 \end{tabular}
  }
 \caption{{\small Summary of linear relationships (Df:23, $\alpha=0.05$) between the expression attributes and human ratings on complexity }}
 \label{tbl:expressionAttributes}
  \end{center}
 \end{table}

    We cast a regression problem that learns a mapping between
    the structural properties of an expression and the degree of human interpretability as reported by our participants.
    We trained a linear model with leave-one-out cross-validation.

    \begin{align*}
      Predicted Human Rating(PHR) = \\
      &0.0854  * \text{Tree Depth} + \\
      &-0.2568 * \text{Number of Blocks} +\\
      &-0.1014 * \text{Avg. Block Size} +\\
      &0.0899 *  \text{Total Size} +\\
      &0.2151
    \end{align*}

    This linear model is highly predictive of the human ratings with respect to tree depth, number and average size of the blocks,
    and the total size (Pearson's R=$ 0.9598$, Df=23, $p < 0.05$). Based on this model, we infer that humans rate longer and nested expressions as more
    difficult to interpret while the existence of small number of compact blocks increase the interpretability.

\section{Discussion}
\label{sect:conclusion}

 In this paper, we investigated the relationships between human perception and automated measures that aim to quantify
interpretability. For visual exploration of high dimensional labeled datasets, we considered two forms of interpretability.
Visual interpretability is concerned with how easy it is to tell the members of different classes apart by looking at 2D scatterplots
of data. We argued that classifier performance and various clustering validity measures from the machine learning literature can
also be used to assess the quality of the views besides the recently proposed visual quality measures. We presented a user study on
four datasets with varying number of classes comparing ten automated measures to human perception. Our results indicated that no
single measure outperforms others on all datasets. While the Dunn Index, C Index and LDA Index might not correlate with human
perception well depending on the dataset, all other measures seem to correlate with human perception to some extent.
However, a linear combination of a subset of the automated measures correlated significantly with human perception across 
all scatterplots as well as individual datasets.

Semantic interpretability is concerned with how easy it is for humans to understand the data transformation functions that project
the original features into lower dimensions. We investigated how humans would rate expressions of varying level of complexity.
We found that up to expression size 7 (up to 4 operands or 4 operators), the participants rated the expressions to be very easy-easy
(rating $\leq$ 0.3). Based on a linear combination of various structural properties of an expression, we inferred that humans rated
longer and nested expressions as more difficult to interpret while the existence of small number of
compact blocks increase the interpretability.

In exploratory analysis of labeled data, simple feature-feature combinations might not always be the best views that reveal the
class structures. Linear or non-linear feature transformation functions might create better 2D views. However, the space of possible
transformation functions is vast. Therefore, automated complexity measures reflecting the human perception closely will
be useful in finding interpretable data transformations.

The automated measures for visual and semantic interpretability of data transformations can be combined in a number of ways in order
to search for good views of data that are also easily understandable in terms of the original attributes. A weighted linear
combination of visual and semantic interpretability measures can be utilized or they can be optimized simultaneously using a
multi-objective optimization scheme.

In conclusion, we state that through investigation of automated measures of visual and semantic interpretability,
we can improve exploratory analysis
by simultaneously presenting data representations to a user that are both easy to visualize and whose axes represent
dimensions that are transparently understood.

\bibliographystyle{abbrv}


\bibliography{references}

\begin{thebibliography}{10}

\bibitem{oliveoil}
Italian olive oils dataset.
\newblock http://www.ggobi.org/book/data/olive.csv.

\bibitem{uci}
Uci machine learning data repository.
\newblock \url{www.ics.uci.edu/~mlearn}.

\bibitem{weka}
Weka machine learning toolkit.
\newblock \url{http://www.cs.waikato.ac.nz/~ml/weka/}.

\bibitem{anthony2005}
L.~Anthony, J.~Yang, and K.~R. Koedinger.
\newblock Evaluation of multimodal input for entering mathematical equations on
  the computer.
\newblock In {\em CHI '05 extended abstracts on Human factors in computing
  systems}, CHI EA '05, pages 1184--1187, New York, NY, USA, 2005. ACM.

\bibitem{awde2008}
A.~Awde, Y.~Bellik, and C.~Tadj.
\newblock Complexity of mathematical expressions in adaptive multimodal
  multimedia system ensuring access to mathematics for visually impaired users.
\newblock {\em International Journal of Computer and Information Science and
  Engineering}, 2:103--115, 2008.

\bibitem{Bellman1961}
R.~Bellman.
\newblock {\em Adaptive Control Processes}.
\newblock Princeton Univ. Press, 1961.

\bibitem{burges2005}
C.~J.~C. Burges.
\newblock Geometric methods for feature extraction and dimensional reduction.
\newblock pages 59--92, 2005.

\bibitem{db1979}
D.~L. Davies and D.~W. Bouldin.
\newblock A cluster separation measure.
\newblock {\em IEEE Transactions on Pattern Analysis and Machine Intelligence},
  PAMI-1(2):224--227, Apr. 1979.

\bibitem{dunn1974}
J.~C. Dunn.
\newblock Well separated clusters and optimal fuzzy-partitions.
\newblock {\em Journal of Cybernetics}, 4:95--104, 1974.

\bibitem{friedman1974}
J.~H. Friedman and J.~W. Tukey.
\newblock A projection pursuit algorithm for exploratory data analysis.
\newblock {\em Computers, IEEE Transactions on}, C-23(9):881--890, 1974.

\bibitem{guyon2006}
I.~Guyon, S.~Gunn, M.~Nikravesh, and L.~Zadeh, editors.
\newblock {\em Feature Extraction, Foundations and Applications}.
\newblock Series Studies in Fuzziness and Soft Computing, Physica-Verlag,
  Springer, 2006.

\bibitem{halkidi2002}
M.~Halkidi, Y.~Batistakis, and M.~Vazirgiannis.
\newblock Cluster validity methods: part i.
\newblock {\em SIGMOD Rec.}, 31:40--45, June 2002.

\bibitem{hand2001}
D.~J. Hand and K.~Yu.
\newblock Idiot's {Bayes---Not} so stupid after all?
\newblock {\em International Statistical Review}, 69(3):385--398, 2001.

\bibitem{hubert1976}
L.~Hubert and J.~Schultz.
\newblock Quadratic assignment as a general data-analysis strategy.
\newblock {\em British Journal of Mathematical and Statistical Psychologie},
  29:190--241, 1976.

\bibitem{elarini2006}
T.~L. Khalid El-Arini, Andrew W.~Moore.
\newblock Autonomous visualization.
\newblock In {\em European Conference on Principles and Practice of Knowledge
  Discovery in Databases (ECML/PKDD 2006)}, Berlin, Germany, September 2006.

\bibitem{kohavi1997}
R.~Kohavi and G.~H. John.
\newblock Wrappers for feature subset selection.
\newblock {\em Artif. Intell.}, 97(1-2):273--324, 1997.

\bibitem{leban2006}
G.~Leban, B.~Zupan, G.~Vidmar, and I.~Bratko.
\newblock Vizrank: Data visualization guided by machine learning.
\newblock {\em Data Mining and Knowledge Discovery}, 13:119--136, 2006.
\newblock 10.1007/s10618-005-0031-5.

\bibitem{lee2005}
E.-K. Lee, D.~Cook, S.~Klinke, and T.~Lumley.
\newblock Projection pursuit for exploratory supervised classification.
\newblock {\em Journal of Computational and Graphical Statistics},
  14(4):831--846, December 2005.

\bibitem{morton1989}
S.~C. Morton.
\newblock {\em Interpretable Projection Pursuit}.
\newblock PhD thesis, Stanford University, 1989.

\bibitem{rendon2011}
E.~Rend\'{o}n, I.~M. Abundez, C.~Gutierrez, S.~D. Zagal, A.~Arizmendi, E.~M.
  Quiroz, and H.~E. Arzate.
\newblock A comparison of internal and external cluster validation indexes.
\newblock In {\em Proceedings of the 2011 American conference on applied
  mathematics and the 5th WSEAS international conference on Computer
  engineering and applications}, AMERICAN-MATH'11/CEA'11, pages 158--163,
  Stevens Point, Wisconsin, USA, 2011. World Scientific and Engineering Academy
  and Society (WSEAS).

\bibitem{rubinstein1997}
Y.~D. Rubinstein and T.~Hastie.
\newblock Discriminative vs informative learning.
\newblock In {\em KDD}, pages 49--53, 1997.

\bibitem{sips2009}
M.~Sips, B.~Neubert, J.~P. Lewis, and P.~Hanrahan.
\newblock Selecting good views of high-dimensional data using class
  consistency.
\newblock {\em Computer Graphics Forum}, 28(3):831--838, 2009.

\bibitem{tatu2009}
A.~Tatu, G.~Albuquerque, M.~Eisemann, J.~Schneidewind, H.~Theisel, M.~Magnor,
  and D.~Keim.
\newblock Combining automated analysis and visualization techniques for
  effective exploration of high-dimensional data.
\newblock In {\em Proceedings of the IEEE Symposium on Visual Analytics Science
  and Technology (IEEE VAST)}, pages 59--66, Atlantic City, New Jersey, USA, 10
  2009.

\bibitem{tatu2010}
A.~Tatu, P.~Bak, E.~Bertini, D.~Keim, and J.~Schneidewind.
\newblock Visual quality metrics and human perception: an initial study on 2d
  projections of large multidimensional data.
\newblock In {\em Proceedings of the International Conference on Advanced
  Visual Interfaces}, AVI '10, pages 49--56, New York, NY, USA, 2010. ACM.

\bibitem{maaten2009}
L.~van~der Maaten, E.~Postma, and H.~van~den Herik.
\newblock Dimensionality reduction: A comparative review.
\newblock Technical report, Tilburg University, TiCC-TR 2009-005, 2009.

\bibitem{wu2007}
X.~Wu, V.~Kumar, J.~Ross~Quinlan, J.~Ghosh, Q.~Yang, H.~Motoda, G.~J.
  McLachlan, A.~Ng, B.~Liu, P.~S. Yu, Z.-H. Zhou, M.~Steinbach, D.~J. Hand, and
  D.~Steinberg.
\newblock Top 10 algorithms in data mining.
\newblock {\em Knowl. Inf. Syst.}, 14:1--37, December 2007.

\bibitem{zupan1994}
J.~Zupan, M.~Novic, X.~Li, and J.~Gasteiger.
\newblock Classification of multicomponent analytical data of olive oils using
  different neural networks.
\newblock {\em Analytica Chimica Acta}, 292(3):219 -- 234, 1994.

\end{thebibliography}
\end{document}